\documentclass[a4paper,11pt]{article}
\usepackage{pos}
\usepackage{subcaption}

\title{New results for thermal interquark bottomonium potentials using NRQCD from the HAL QCD method }

\ShortTitle{Thermal interquark bottomonium potentials}
\author*[a]{Thomas Spriggs}
\author[a]{Chris Allton}
\author[a]{Timothy Burns}
\author[b]{Seyong Kim}

\affiliation[a]{Department of Physics, Swansea University, Swansea SA2 8PP, United Kingdom}

\affiliation[b]{Department of Physics, Sejong University, Seoul 143-747, Korea}

\emailAdd{\{t.spriggs.996870,c.allton,t.burns\}@swansea.ac.uk}
\emailAdd{skim@sejong.ac.kr}

\abstract{
We report progress in the calculation of the thermal interquark potential of bottomonium using the HAL QCD method applied to bottom quarks in the non-relativistic approximation (i.e. NRQCD). We exploit the fast Fourier transform algorithm, using a momentum space representation, to efficiently calculate NRQCD correlation functions of non-local mesonic S-wave states, and thus obtain the potential for temperatures in both the hadronic and plasma phases. This work was performed on our anisotropic 2+1 flavour ``Generation 2" FASTSUM ensembles.
}

\FullConference{%
  The 39th International Symposium on Lattice Field Theory, LATTICE2022
  8th–13th August, 2022
  Bonn, Germany
}

\usepackage{tikz}
\usetikzlibrary{arrows}

\begin{document}
\maketitle

\section{Introduction}

The interquark potential of quarkonia was one of the first quantities studied in the quest for a deeper understanding of the nature of the strong interaction.  Pioneering studies include \cite{Eichten:1978tg} where the Cornell potential was used to calculate the spectrum of charmonium states using a quantum mechanical formalism.  In thermal QCD, the temperature dependence of the interquark potential results in quarkonium states melting at different temperatures \cite{Matsui:1986dk}.  These considerations strongly motivate a study of the thermal behaviour of the quarkonia interquark potential.

Slowly moving heavy quarks, interacting via QCD, can be studied using non-relativistic QCD (NRQCD) which allows significant benefits. For example, NRQCD calculations of bottomonia are typically accurate at the percent level or less and is an excellent ground for quantitative tests. In this work we use NRQCD to determine the interquark potential in bottomonia using the HAL QCD approach \cite{Ishii:2006ec}: Correlation functions of bottomonia operators are studied where the quark and antiquark are spatially separated, and this allows an access to the Nambu-Bethe-Salpeter wavefunction in the quarkonium rest frame. Using this wavefunction in the Schr\"odinger equation leads to the interquark potential. We find indications of the weakening of the potential as the temperature increases, as expected. This work is a continuation of the work in \cite{Spriggs:2021ieo} and extends previous studies of the interquark potential by the FASTSUM Collaboration in the charmonium system \cite{Evans:2013yva,Allton:2015ora}. Other work in this area includes \cite{Larsen:2020rjk}.

\section{NRQCD and lattice setup}

NRQCD is an effective theory with a power counting in the heavy quark velocity, $v$. In this theory, the heavy quark and antiquark fields decouple and so virtual heavy quark-antiquark loops cannot form. The NRQCD quark propagator is calculated via an initial value problem, rather than via a boundary value problem (as is the case for relativistic quarks). NRQCD is particularly amenable for lattice simulations because NRQCD quarkonium correlation functions do not have ``backward movers'' which means the full extent of the lattice in the temporal direction can be used in the analysis.

Our NRQCD formulation incorporates both ${\cal O}(v^4)$ and the leading spin-dependent corrections. The $b$-quark mass is tuned by setting the ``kinetic'' mass (i.e. from the dispersion relation) of the spin-averaged $1S$ states to its experimental value. Full details of our NRQCD setup appear in \cite{Aarts:2014cda}.

All our results were obtained using our FASTSUM $N_f=2+1$ flavour ``Generation 2'' ensembles which have the parameters listed in Table \ref{tab:fastsum_details}.

\begin{table}[h!]
    \centering
        \begin{tabular}{c||c|c|c|c||c|c|c}
        $N_\tau$ & 16 & 20 & 24 & 28 & 32 & 36 & 40 \\
        \hline
        T [MeV] & 352 & 281 & 235 & 201 & 176 & 156 & 141 \\
        \hline
        $N_{\text{configurations}}$ & 1050 & 950 & 1000 & 1000 & 1000 & 500 & 500 
    \end{tabular}
    \caption{An overview of the FASTSUM Generation 2 correlation
      functions used in this work. Lattice volumes are $(24 a_s)^3\times (N_\tau
      a_\tau)$ with $a_s = 0.1227(8)$fm and $a_\tau =
      35.1(2)$am. For these ensembles with a pion mass of $M_\pi = 384(4)$MeV,
      the pseudo-critical temperature T$_{\rm pc} = 181(1)$MeV \cite{Aarts:2019hrg}. }
    \label{tab:fastsum_details}
\end{table}

\section{Method}

\subsection{The HAL QCD method}\label{sec:hal}
To calculate the potential between two quarks in a bottomonium - the interquark potential, $V(r)$ - we use the method from the HAL QCD collaboration \cite{Ishii:2006ec}. In brief, this method uses the point-split correlation function and the time independent Schr\"odinger equation to calculate the interquark potential.

\begin{figure}[h]
\centering
\begin{tikzpicture}
% Lines 
\draw (3,0) -- node {\tikz \draw[-triangle 90](0,0) -- +(-.1,0);} (0,0);
\draw (0,0) arc [
				start angle = 0,
				end angle = 90,
				x radius=-3cm,
				y radius=1cm
				];
\draw [dashed] (3,1) -- (3,0);
\draw[-triangle 90](1.52,0.865) -- +(.1,.02);

% Nodes
\filldraw[black] (0,0) circle (2pt) node[anchor=north]{(\textbf{0},0)};
\filldraw[black] (3,0) circle (2pt) node[anchor=north]{(\textbf{x},$\tau$)};
\filldraw[black] (3,1) circle (2pt) node[anchor=south]{(\textbf{x}+\textbf{r},$\tau$)};

% Text
\filldraw[black] (-1,0) node[anchor=south]{$J^{\dagger}_\Gamma(0;\textbf{0})$};
\filldraw[black] (4,0) node[anchor=south]{$J_\Gamma(\textbf{x}, \tau;\textbf{r})$};

\filldraw[black] (0,2.2) node[anchor=north]{Source};
\filldraw[black] (3,2.2) node[anchor=north]{Sink};

\filldraw[black] (1.5,-0.1) node[anchor=north]{$\bar{b}$};
\filldraw[black] (1.5,1) node[anchor=south]{$b$};

\end{tikzpicture}
\caption{A representation of the point-split correlation function, as defined in \eqref{eq:point-split-corr}}
\label{fig:corr_drawing}
\end{figure}
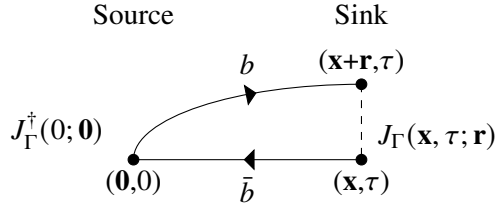

The point-split correlation function is defined by
\begin{equation}
C_{\Gamma}(\textbf{r},\tau) = \sum_{\textbf{x}} \langle J_\Gamma(\textbf{x}, \tau;\textbf{r}) J^{\dagger}_\Gamma(0;\textbf{0}) \rangle,
\label{eq:point-split-corr}
\end{equation}
where the non-local mesonic operators are defined
\begin{equation}
J_\Gamma(x;\textbf{r}) = \bar{q}(x)\Gamma U(x,x+\textbf{r})q(x+\textbf{r}).
\label{eq:J}
\end{equation}
The quark and antiquark fields, $q$ and $\bar{q}$, are separated in space by $\textbf{r}$. The gauge field $U(x,x+\textbf{r})$ is required to ensure gauge invariance and
$\Gamma$ signifies the channel being considered; in this work we consider vector and pseudoscalar S-wave states.
The correlator in \eqref{eq:point-split-corr} is depicted in Figure \ref{fig:corr_drawing}.

%As usual, the correlation function can be expressed as a sum over eignstates of the Hamiltonian,
%\begin{equation}
%C_\Gamma(\textbf{r},\tau) = \sum_j \frac{\psi^*_j(\textbf{0})\psi_j(\textbf{r})}{2E_j} e^{-E_j\tau},
%\label{eq:C_estate}
%\end{equation}
%where $E_j$ is the energy of a given state $j$, and $\psi_j(\textbf{r})$ is the Nambu-Bethe–Salpeter wavefunction. We will denote an unnormalised wavefunction $\Psi_j(\textbf{r}) = \dfrac{\psi^*_j(\textbf{0})\psi_j(\textbf{r})}{2E_j}$ for brevity. 

As usual, the correlation function can be expressed as a sum over eigenstates of the Hamiltonian,
\begin{equation}
C_\Gamma(\textbf{r},\tau) = \sum_j \Psi_j(\textbf{r}) e^{-E_j\tau},
\label{eq:C_estate}
\end{equation}
where $E_j$ is the energy of a given state $j$, and the 
unnormalised wavefunction 
\begin{equation}\Psi_j(\textbf{r}) = \dfrac{\psi^*_j(\textbf{0})\psi_j(\textbf{r})}{2E_j}\end{equation}
is defined in terms of the Nambu-Bethe–Salpeter wavefunction $\psi_j(\textbf{r})$. 

We introduce the time-independent Schr\"odinger equation,
\begin{equation}
\left( -\frac{\nabla^2_r}{2 \mu} + V_\Gamma \left(r\right) \right)\Psi_j\left(r\right) = E_j\Psi_j\left(r\right),
\end{equation}
where $V_\Gamma(r)$ is the potential for the channel $\Gamma$ and $\mu$ is the reduced quark mass. We apply the Schr\"odinger equation to the point-split correlation function in \eqref{eq:C_estate} through the following steps
\begin{equation}
\begin{split}
-\frac{\partial C_\Gamma(\textbf{r},\tau)}{\partial \tau} = \sum_j E_j\Psi_j(\textbf{r})e^{-E_j\tau} &= \sum_j \left(-\frac{\nabla^2_r}{2 \mu} + V_\Gamma \left(r\right) \right)\Psi_j\left(r\right)e^{-E_j\tau}\\
& = \left(-\frac{\nabla^2_r}{2 \mu} + V_\Gamma \left(r\right) \right)C_\Gamma(\textbf{r},\tau).
\end{split}
\end{equation}
This yields the form of the interquark potential for a given channel, $V_\Gamma$, as
\begin{equation}
V_\Gamma(r) = \frac{1}{C_\Gamma(\textbf{r},\tau)}\left(\frac{\nabla^2_r}{2 \mu} - \frac{\partial}{\partial \tau} \right) C_\Gamma(\textbf{r},\tau).
\label{eq:V_c}
\end{equation}
Note that in the continuum limit, we expect the potential to be function of $r=|\textbf{r}|$.
There is explicit time dependency in this form for the potential, and this will be studied in Section \ref{sec:time_dep}.
Section \ref{sec:quark_mass} will discuss how the reduced quark mass, $\mu$, is set.

It is convenient to define the central potential, $V_{\text{C}}$, obtained via the usual spin-average \cite{Godfrey:1985xj}
\begin{equation}
V_{\text{C}} = \frac{1}{4}V_{\text{Pseudo Scalar}} + \frac{3}{4}V_{\text{Vector}}.
\end{equation}
\color{black}

\subsection{Using momentum space to reformulate the calculation}
\label{sec:mom_space}
This work is a continuation of \cite{Spriggs:2021ieo} where more detail about the HAL QCD method can be found. We build upon  \cite{Spriggs:2021ieo} by using an efficient computation of the point-split correlation function, $C_{\Gamma}(\textbf{r},\tau)$.

For each $\tau$, a direct calculation of \eqref{eq:point-split-corr} requires a loop over all lattice sites $\textbf{x}$ for each value of $\textbf{r}$ which is an expensive operation scaling as $\mathcal{O}(\mathcal{V}^2)$ where $\mathcal{V}$ is the spatial volume.
What follows is a method to reduce the cost of this computation by introducing a momentum space representation for the propagator and correlation function, see the Appendix of \cite{Allton:2015ora}.

We introduce quark propagators, $D^{-1}(x;y)$, by Wick contracting the quark fields in the point-split correlation function, \eqref{eq:point-split-corr},
\begin{equation}
C_{\Gamma}(\textbf{r},\tau) = - \sum_{\textbf{x}} \langle D^{-1}(\textbf{x}+\textbf{r}, \tau; \textbf{0},0) \Gamma \gamma_5 \left( D^{-1}(\textbf{x}, \tau; \textbf{0},0) \right) ^{\dagger} \gamma_5 \Gamma^{\dagger} \rangle.
\label{eq:point-split-ito-propagator}
\end{equation}

Note that we have gauge fixed our configurations to the Coulomb gauge, and have replaced the gauge connection, $U(x,x+\textbf{r})$ in \eqref{eq:J} by unity. 
We now implicitly define the corresponding momentum space quark propagator via
\begin{equation}
D^{-1}(\textbf{y},\tau;\textbf{0},0) = \frac{1}{V}\sum_{\textbf{p}}\tilde{D}^{-1}(\textbf{p},\tau)e^{i\textbf{y}\cdot\textbf{p}},
\label{eq:ft_of_prop}
\end{equation}
in terms of the 3-momentum, $\textbf{p}$, which is conjugate to the position $\textbf{y}$.
Introducing this momentum-space quark propagator into \eqref{eq:point-split-ito-propagator} yields
\begin{equation}
C_{\Gamma}(\textbf{r},\tau) = \frac{1}{V}\sum_{\textbf{p}}\langle \tilde{D}^{-1}(\textbf{p},\tau) \Gamma \gamma_5 \tilde{D}^{-1}(-\textbf{p},\tau)\gamma_5 \Gamma^{\dagger}\rangle e^{i\textbf{p} \cdot \textbf{r}},
\end{equation}
which we will use to implicitly define the momentum-space correlator, $\tilde{C}_\Gamma (\textbf{p},\tau)$, \textit{i.e.}
\begin{equation}
C_\Gamma(\textbf{r},\tau) = \frac{1}{V}\sum_{\textbf{p}} \tilde{C}_\Gamma (\textbf{p},\tau) e^{i\textbf{p} \cdot \textbf{r}}.
\label{eq:corr-mom-space}
\end{equation}

%\subsubsection{The fast Fourier transform} \label{sec:FFT}
We note that once we have calculated $\tilde{C}_\Gamma (\textbf{p},\tau)$, we can determine the desired correlator $C_\Gamma(\textbf{r},\tau)$ for any $\textbf{r}$ using \eqref{eq:corr-mom-space}.

At first sight, the conversion to momentum space does not produce any savings, because the calculation of $C_\Gamma$ and $\tilde{D}^{-1}$, defined via \eqref{eq:ft_of_prop} and \eqref{eq:corr-mom-space}, are both $\mathcal{O}(\mathcal{V}^2)$ in the number of operations, i.e. the same as the direct method.
However both \eqref{eq:ft_of_prop} and \eqref{eq:corr-mom-space} are Fourier transforms, and so significant speed-up for these steps can be achieved using the fast Fourier transform (FFT) algorithm which scales as $\mathcal{O}(\mathcal{V}\log \mathcal{V})$.

\section{Results}
For better comparison with \cite{Spriggs:2021ieo}, and as progress towards the treatment of $C_\Gamma(\textbf{r},\tau)$ for all $\textbf{r}$, we consider here only the on-axis $\textbf{r}$ data. Extensions to this will be discussed in Section \ref{sec:conclusion_and_further_work}.

\subsection{Time dependence}
\label{sec:time_dep}
\begin{figure}
\centering
\begin{tabular}{cc}
\includegraphics[width=75mm]{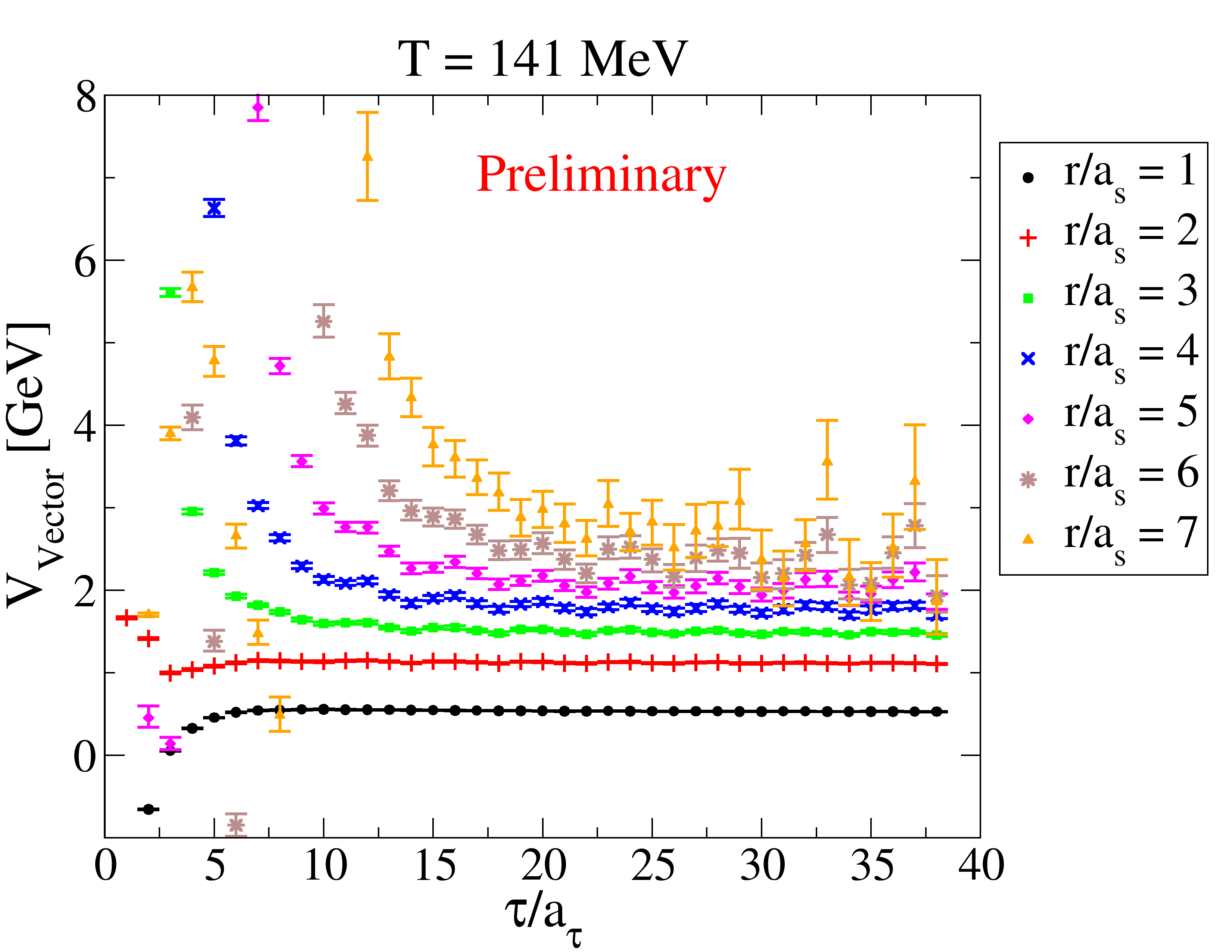} &
\includegraphics[width=75mm]{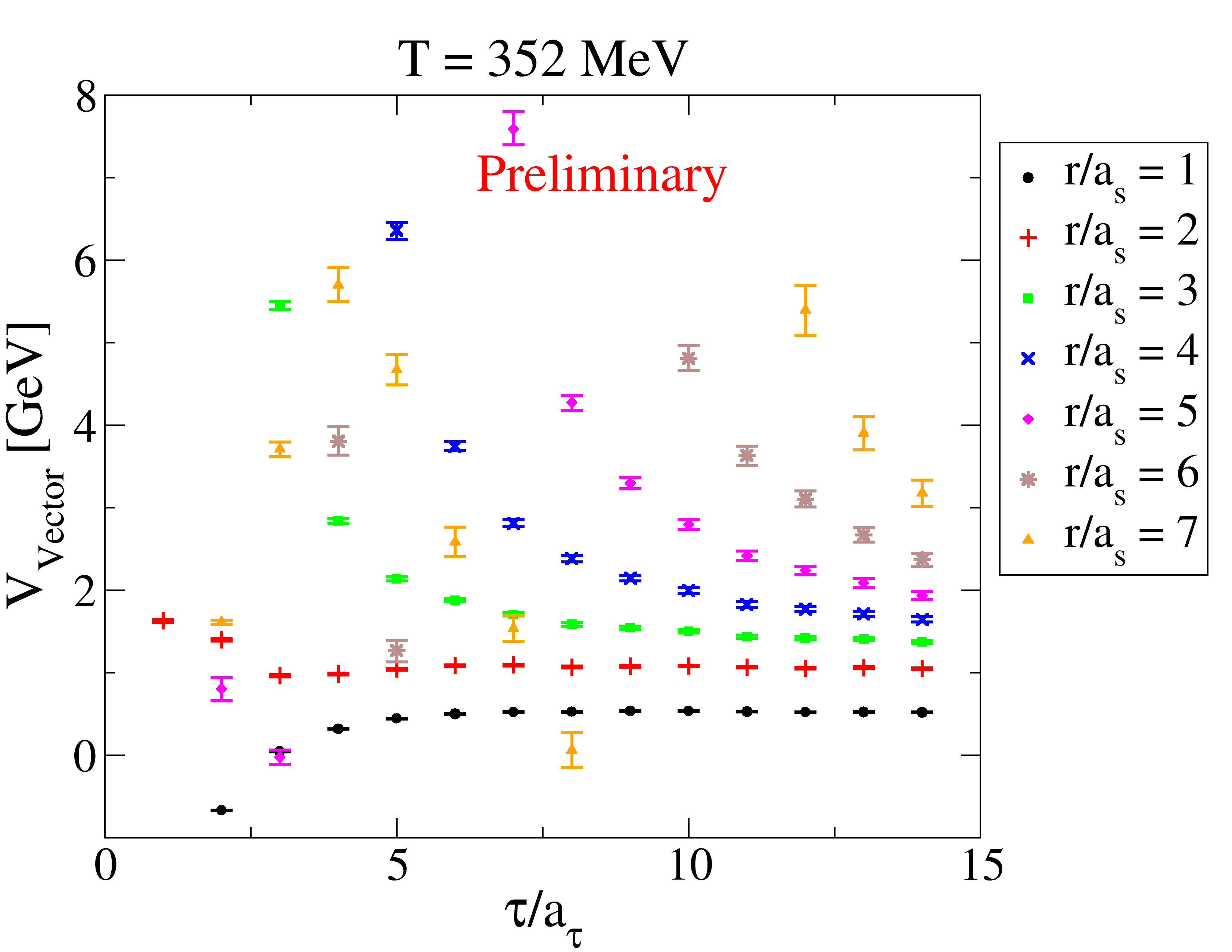} \\
\end{tabular}
\caption{Time dependence in the potential restricting the range of $r$ that we can consider valid. Shown for two temperatures using the vector channel as an example.}
\label{fig:time_dependence}
\end{figure}

The potential is defined in \eqref{eq:V_c} where there is an apparent explicit dependence on time, $\tau$, from the correlation function. In Figure \ref{fig:time_dependence}, the potential, $V_\text{Vector}$, from \eqref{eq:V_c} is plotted  against $\tau$ for a variety of distances $\textbf{r}$ for our two extreme temperatures, $T=141$ and $352$ MeV. We can see a clear $\tau$ dependence for small $\tau$ which increases with $\textbf{r}$. However, for various ranges of $\tau$ and $\textbf{r}$ there are clear plateau. 

In addition, we note that we would like to uncover temperature effects in the potential. The most accurate way of doing this is to compare different temperatures' potentials obtained with the {\em same} time window to avoid contamination by systematic artefacts.

Based on these considerations, we restrict the range of $r$ and $\tau$ used in the determination of the potential to those listed in Table \ref{tab:time_windows}. Notice that in selecting a time window, there is a trade-off between the ranges of $r$ and $T$ for which the potential can be extracted: larger time windows give access to a larger range of $r$, but over a smaller range of $T$.

\begin{figure}
\centering
\begin{tabular}{cc}
\includegraphics[width=75mm]{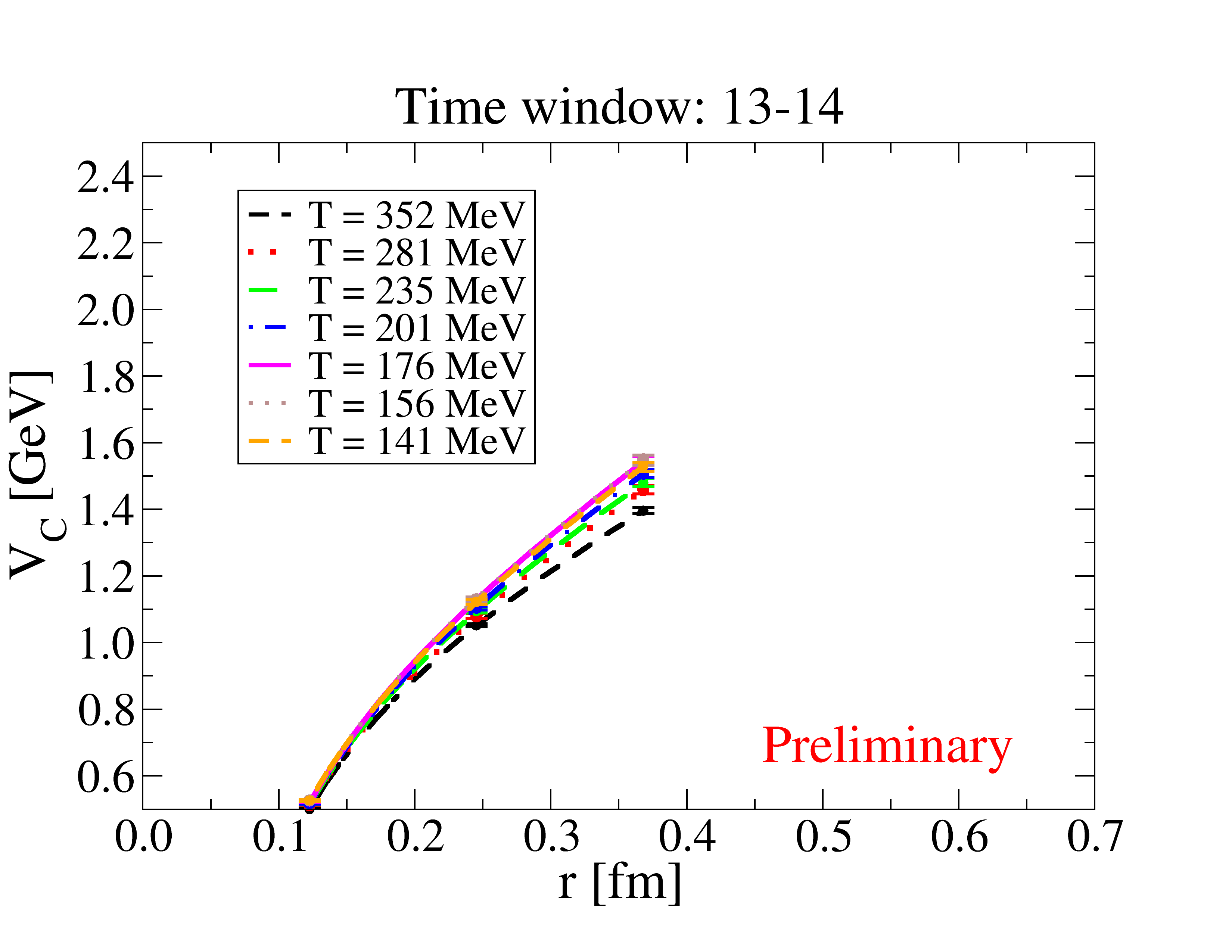} &
\includegraphics[width=75mm]{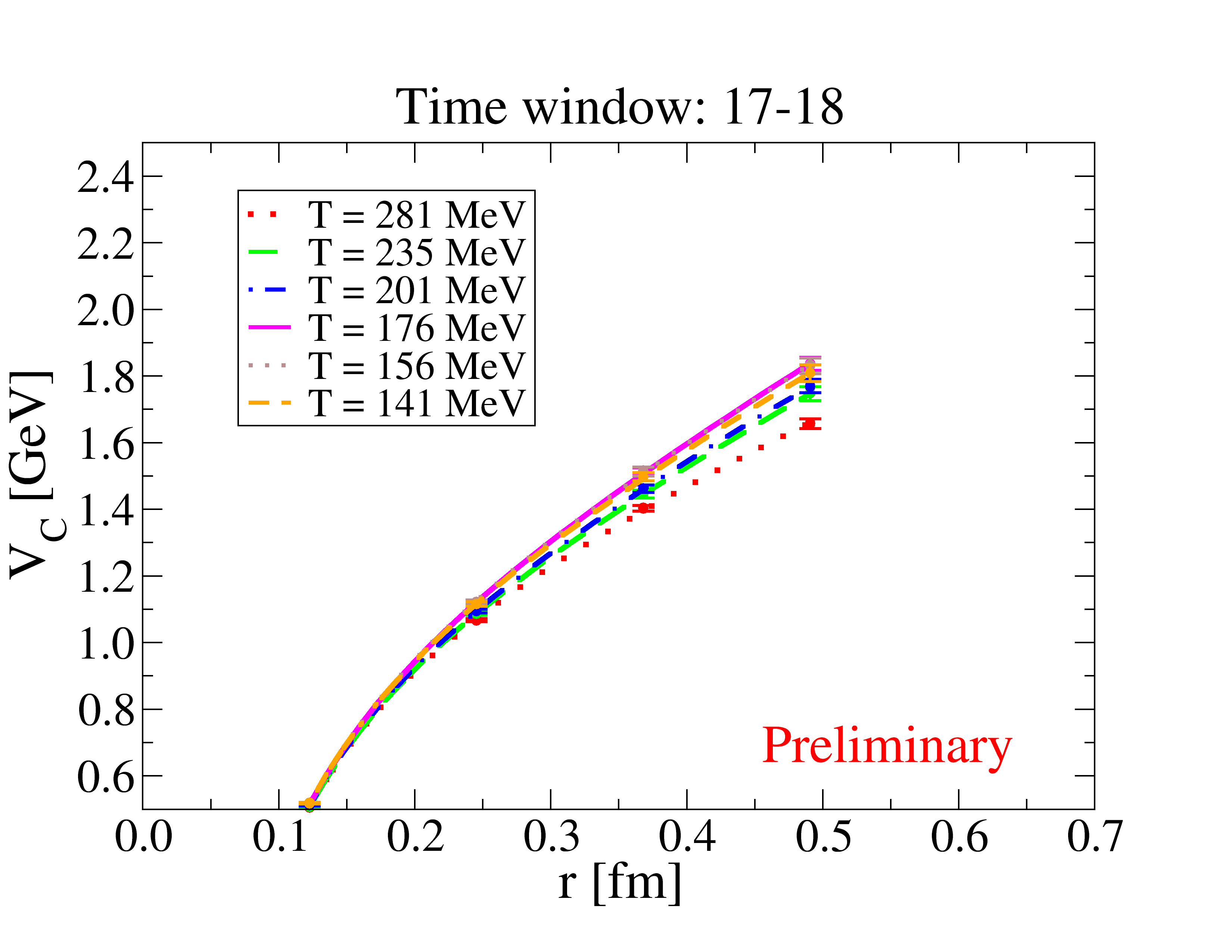} \\
\includegraphics[width=75mm]{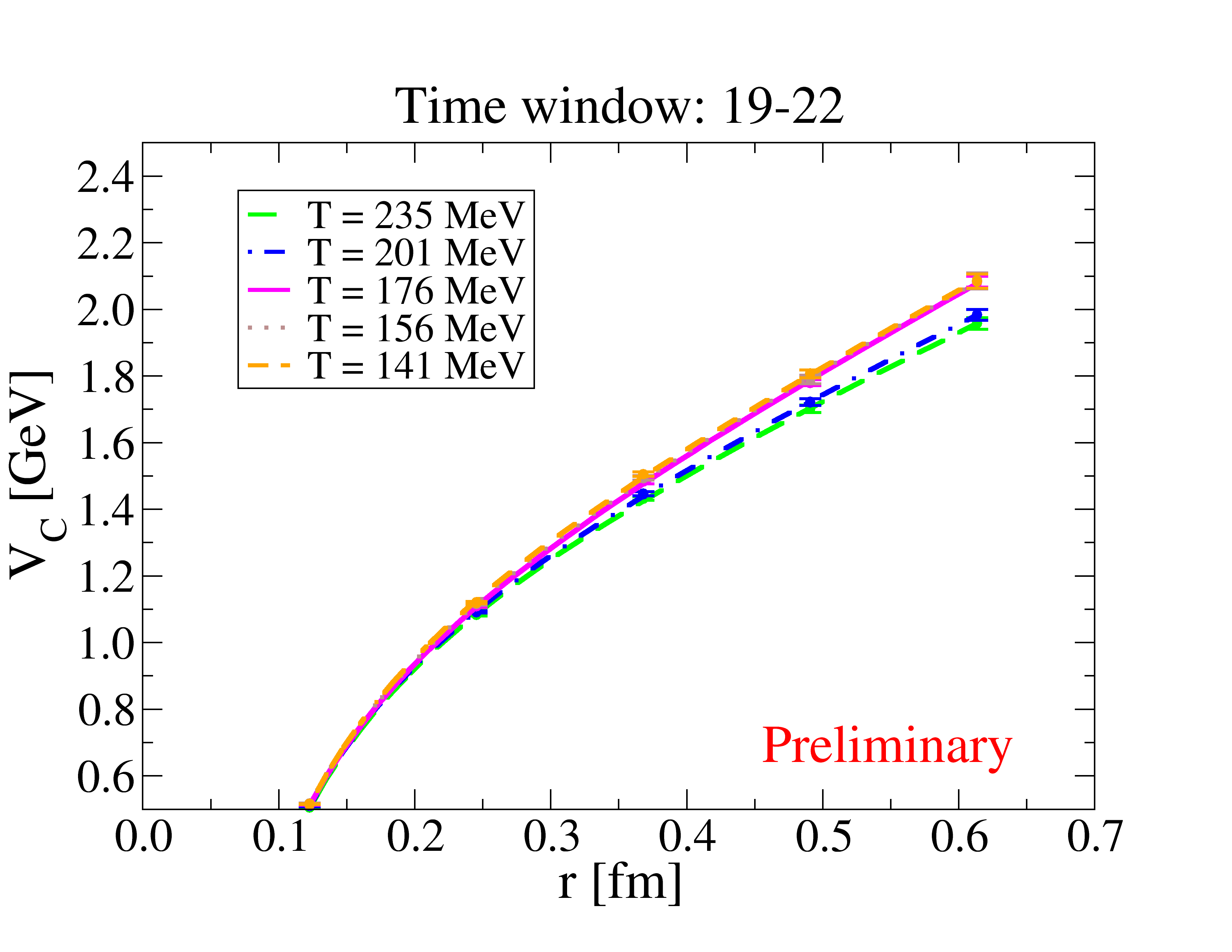} &
\includegraphics[width=75mm]{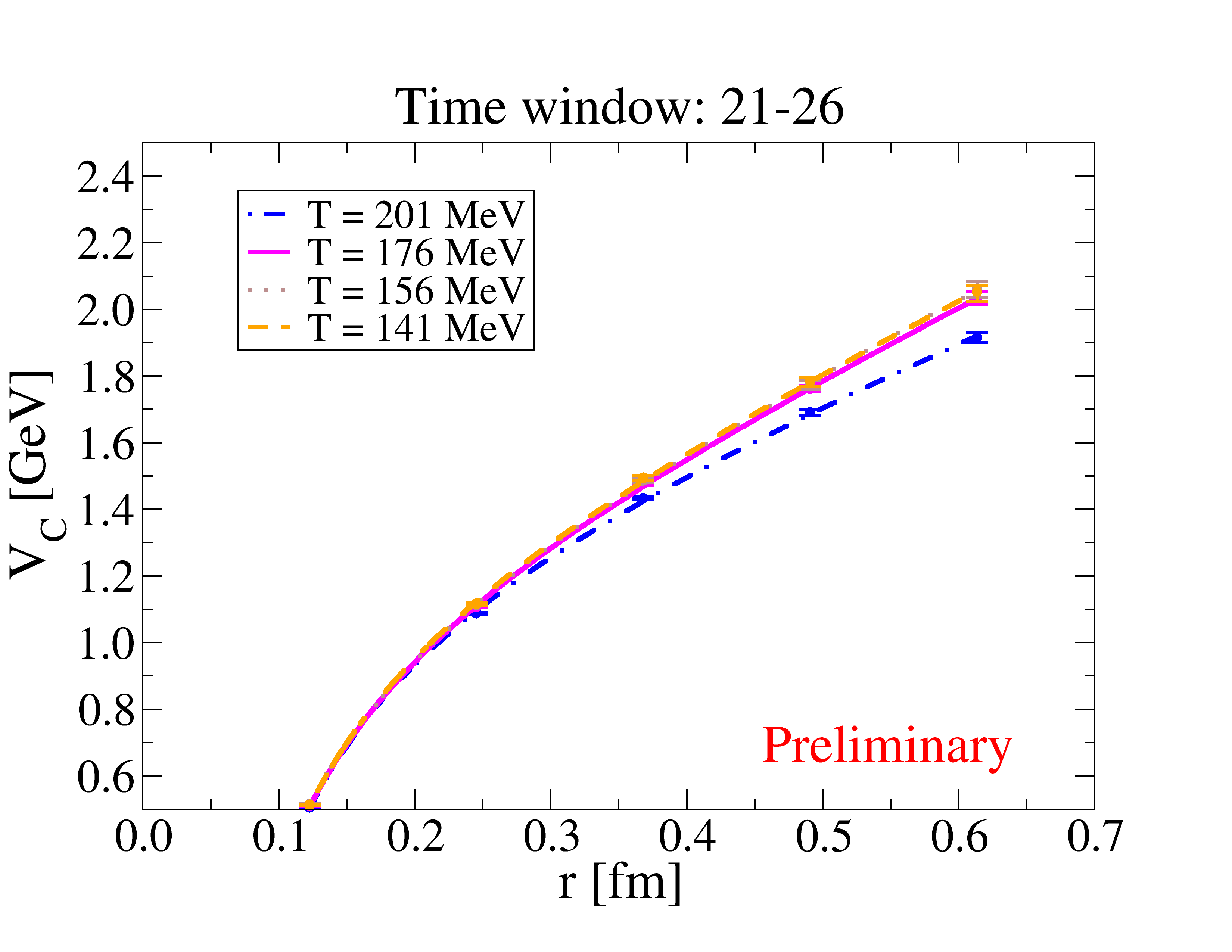} \\
\end{tabular}
\caption{The central potential calculated from \eqref{eq:V_c} (points), overlaid with a fit of these data to the Cornell potential \eqref{eq:cornell} (curves). Each plot contains all temperatures and $\textbf{r}$ ranges listed in Table \ref{tab:time_windows}.}
\label{fig:results}
\end{figure}

In Figure \ref{fig:results} we show four determinations of the central potential, corresponding to the first four time windows identified in Table \ref{tab:time_windows}. In each plot we show the potentials for several temperatures, and since these have been obtained by averaging over the same range of $\tau$, the temperature dependence can be ascribed to temperature effects, rather than fitting artefacts. We find that the potential consistently flattens as the temperature increases above $T_{\text{pc}}$, as expected. There is little thermal variation in the potential for $T \lessapprox T_\text{pc}$.

In Figure \ref{fig:results}, the error bars show statistical errors only. The curves are fits to the Cornell potential, which will be discussed in Section \ref{sec:cornell}.

% Based on these considerations, we restrict the range of $r$ and $\tau$ used in the determination of the potential to those listed in Table \ref{tab:time_windows}. Note that the same range of $\tau$ is used for a variety of temperatures, and so any variation in the calculated potential can be ascribed to temperature effects, rather than fitting artefacts.

%The results are shown in Figure \ref{fig:results} with the error bars showing statistical errors only. Plotted are the calculated central potentials obtained by averaging the potential over the $\tau$-range listed in Table \ref{tab:time_windows} for a variety of temperatures. Fits to the Cornell potential, which will be discussed in Section \ref{sec:cornell}, are also shown. This figure shows that the potential consistently flattens as the temperature increases above $T_{\text{pc}}$, as expected. There is little thermal variation in the potential for $T \lessapprox T_\text{pc}$.

\begin{table}
\centering
\begin{tabular}{c|c|c|c}
Time window [$a_\tau$] & $r$ range [$a_s$] & $r$ range [fm] & Temperatures [MeV]\\
\hline
13-14 & 1-3 & 0.12-0.37 & 352-141\\
17-18 & 1-4 & 0.12-0.49 & 281-141\\
19-22 & 1-5 & 0.12-0.61 & 235-141\\
21-26 & 1-5 & 0.12-0.61 & 201-141\\
\hline
24-30 & 1-6 & 0.12-0.74 & 176-141\\
24-33 & 1-6 & 0.12-0.74 & 156-141\\
%24-38 & 1-6 & 0.12-0.74 & 141\\
\end{tabular}
\caption{Range of displacements and temperatures allowed to best approximate time independence in $V(r,\tau)$. Note that $T_{\text{pc}}=181$ MeV and thus the time windows below the solid line do not span this pseudocritical temperature.}
\label{tab:time_windows}
\end{table}

\subsection{Quark mass dependence} \label{sec:quark_mass}

Equation \eqref{eq:V_c} contains the reduced quark mass, $\mu$, which needs to be defined. In \cite{Larsen:2020rjk}, the 1S and 2S states were used to determine the bottom quark mass $m_b$, and thus the reduced quark mass. In our simulations we do not have access to the 2S state. We instead use the simple argument: $\mu \equiv \frac{1}{2}m_b \approx \frac{1}{2} M_\Upsilon$, with $M_\Upsilon$ from \cite{Workman:2022ynf}.
We have tested the sensitivity of the potential on the quark mass and found that the variation (within sensible $\mu$ ranges) is minimal.

\subsection{Cornell potential fits}
\label{sec:cornell}
The Cornell potential \cite{Eichten:1979ms} is a phenomenological description of a confining potential applicable to heavy quarks in QCD and is given by
\begin{equation}
V(r) = -\frac{\alpha}{r} + \sigma r + D.
\label{eq:cornell}
\end{equation}
Fits using \eqref{eq:cornell} to our potential data are shown as solid curves in Figure \ref{fig:results}. As can be seen these reproduce the data well.
When the string tension, $\sigma$, in the Cornell potential is zero, this implies a deconfined potential. 
In all cases above $T_{\text{pc}}$, we find that $\sigma$ decreases with increasing temperature, confirming the expected thermal behaviour in the bottomonium system. Below $T_{\text{pc}}$ the string tension does not change within statistical errors. 

\section{Conclusion}\label{sec:conclusion_and_further_work}

The temperature dependence of the central interquark potential in the bottomonium system using NRQCD quarks was explored.
This work was an extension of \cite{Spriggs:2021ieo} and use a momentum-space approach which can improve the efficiency of the calculation.
Clear thermal effects in this potential were observed using a method which decoupled systematic ``time window'' artefacts from physical, thermal effects. A systematic flattening of the potential with increasing temperature above $T_{\text{pc}}$ was observed, with no statistically significant variation in the potential for temperatures below $T_{\text{pc}}$.

This work will be extended in a number of directions.
The potential will be calculated at all possible spatial separations, $\textbf{r}$, rather than just the on-axis values used here, and channels beyond the pseudoscalar and vector S-wave states will be included.
Also, a more robust definition of the reduced quark mass will be developed.
Finally, a direct comparison will be made between these bottomonium results and those obtained for the charmonium potential using the same ensembles in \cite{Allton:2015ora}.

\acknowledgments 

% Chris updated this 25/11/21

This work is supported by STFC grant ST/T000813/1.
SK is supported by the National Research Foundation of Korea under grant NRF-2021R1A2C1092701 and grant NRF-2021K1A3A1A16096820, funded by the Korean government (MEST).
%BP has been supported by a Swansea University Research Excellence Scholarship (SURES).
This work used the DiRAC Extreme Scaling service at the University of Edinburgh, operated by the Edinburgh Parallel Computing Centre and the DiRAC Data Intensive service operated by the University of Leicester IT Services on behalf of the STFC DiRAC HPC Facility (www.dirac.ac.uk). This equipment was funded by BEIS capital funding via STFC capital grants ST/R00238X/1, ST/K000373/1 and ST/R002363/1 and STFC DiRAC Operations grants ST/R001006/1 and ST/R001014/1. DiRAC is part of the UK National e-Infrastructure.
This work was performed using PRACE resources at Cineca (Italy), CEA (France) and Stuttgart (Germany) via grants 2015133079, 2018194714, 2019214714 and 2020214714.
We acknowledge the support of the Swansea Academy for Advanced Computing, the Supercomputing Wales project, which is part-funded by the European Regional Development Fund (ERDF) via Welsh Government, and the University of Southern Denmark and ICHEC, Ireland for use of computing facilities.
We are grateful to the Hadron Spectrum Collaboration for the use of their zero temperature ensemble.

\bibliographystyle{JHEP_arxiv}
\bibliography{ref}

\end{document}